\documentclass[10pt]{iopart}

\usepackage{graphicx}
\usepackage{color}
\usepackage{bm}
\usepackage{cite}
\usepackage{multirow}
\usepackage{hyperref}

\definecolor{green}{rgb}{0,0.6,0.1}

\begin{document}

\title{Helping restricted Boltzmann machines with quantum-state representation by restoring symmetry}

\author{Yusuke Nomura}

\address{RIKEN Center for Emergent Matter Science, 2-1 Hirosawa, Wako, Saitama 351-0198, Japan}
\ead{yusuke.nomura@riken.jp}
\vspace{10pt}
\begin{indented}
\item[]September 2020
\end{indented}

\begin{abstract}
The variational wave functions based on neural networks have recently started to be recognized as a powerful ansatz to represent quantum many-body states accurately.  
In order to show the usefulness of the method among all available numerical methods, it is imperative to investigate the performance in challenging many-body problems for which the exact solutions are not available. 
Here, we construct a variational wave function with one of the simplest neural networks, the restricted Boltzmann machine (RBM), and apply it to a fundamental but unsolved quantum spin Hamiltonian, the two-dimensional $J_1$-$J_2$ Heisenberg model on the square lattice. 
We supplement the RBM wave function with quantum-number projections, which restores the symmetry of the wave function and makes it possible to calculate excited states. 
Then, we perform a systematic investigation of the performance of the RBM. 
We show that, with the help of the symmetry, the RBM wave function achieves state-of-the-art accuracy both in ground-state and excited-state calculations.
The study shows a practical guideline on how we achieve accuracy in a controlled manner.  
\end{abstract}

%
%
%
%
\ioptwocol

\section{Introduction}

\label{Sec_intro}

Quantum many-body systems are a source of various fascinating phenomena.
For example, interacting spin systems show magnetism, and such magnetic materials are indispensable in our daily life. 
It is a longstanding challenge in physics to understand the property of quantum many-body systems.

One of the most powerful numerical methods is the quantum Monte Carlo (QMC) method~\cite{QMC_note}. 
In the QMC method, the multi-dimensional integrals, which appear in many-body problems, are evaluated using the Monte Carlo method.
With the QMC method, one can obtain numerically exact results for various physical quantities such as the total energy and correlation functions. 
However, the QMC method is not applicable when the negative-sign problem becomes severe. 

Another useful method is based on variational wave functions, 
in which the exact wave functions are approximated by some wave function form determined by variational parameters. 
This approach can be applied to, for example, fermion systems for which the QMC method suffers from the sign problem. 
A main difficulty in the variational wave function approach is how to prepare a powerful variational wave function. 
So far, various wave function methods such as the variational Monte Carlo (VMC) method~\cite{McMillan_1965,Ceperley_1977,Yokoyama_1987_1,Yokoyama_1987_2},
the density matrix renormalization group (DMRG)~\cite{White_1992,White_1993},
and 
tensor network methods~\cite{Verstraete_2008,Orus_2014}, have been developed. 

\begin{figure}[tbp]
\vspace{0cm}
\begin{center}
\includegraphics[width=0.35\textwidth]{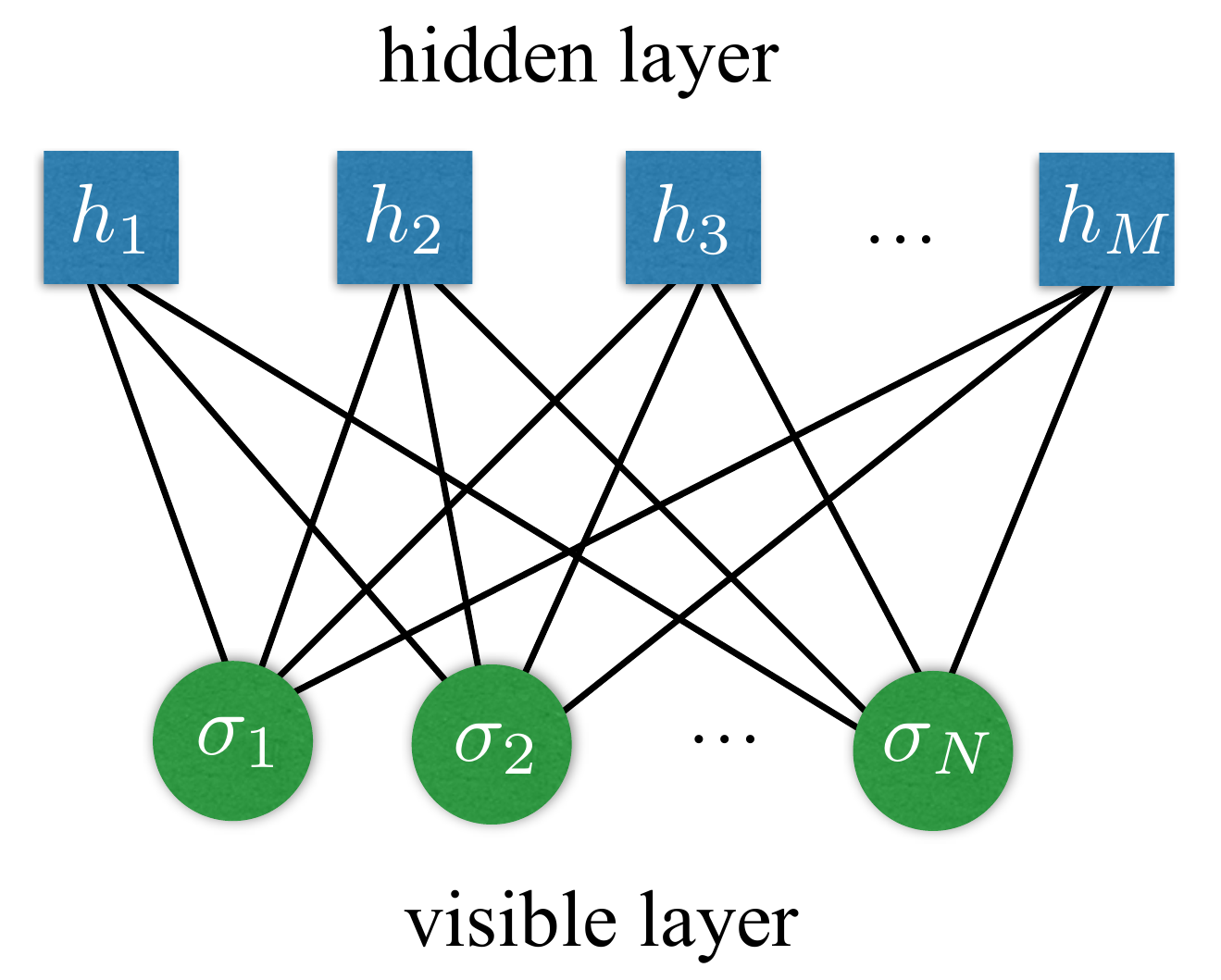}
\caption{
Structure of restricted Boltzmann machine (RBM) with $N$ visible units $\sigma_i = \pm 1$ ($i=1 ,\ldots, N$) and $M$ hidden units $h_k = \pm 1$ ($k=1 ,\ldots,M$). 
The solid lines show the interaction between visible and hidden units. 
}
\label{Fig_structure_RBM}
\end{center}
\end{figure}

Recently, Carleo and Troyer~\cite{Carleo_2017} have proposed a different type of variational wave function based on the restricted Boltzmann machine (RBM)~\cite{Smolensky_1986}. 
The RBM is a neural network composed of two layers (Fig.~\ref{Fig_structure_RBM}), and it approximates the probability distribution function over visible-unit configurations. 
By identifying the states of the physical degrees of freedom with those of visible units and interpreting the wave function amplitude as generalized probability extended to complex numbers, the many-body wave functions can be expressed by the RBM. 
Then, the RBM wave function can be considered as a particular form of the variational wave function in the VMC method, where the parameters in the RBM network play a role as variational parameters.
The RBM allows flexible representation of various quantum states~\cite{Carleo_2019}, including states with volume-law entanglement entropy~\cite{Deng_2017,Chen_2018}.

The neural-network wave functions, including the RBM wave function, were first introduced in the quantum spin systems without geometrical frustration~\cite{Carleo_2017}. 
Later, the applicability has been extended to frustrated spin models~\cite{Cai_2018,Liang_2018,Choo_2019,Ferrari_2019,Westerhout_2020,Szabo_2020,Nomura_arXiv},
itinerant boson systems~\cite{Saito_2017,Saito_2018},
fermion systems~\cite{Cai_2018,Nomura_2017,Luo_2019,Han_2019,Choo_2020,Pfau_2020,Hermann_2020},
fermion-boson coupled systems~\cite{Nomura_2020},
topological states~\cite{Deng_2017,Deng_2017_2,Glasser_2018,Clark_2018,Sirui_2019,Kaubruegger_2018,Huang_arXiv},
excited states~\cite{Choo_2018,Nomura_2020,Nomura_arXiv,Vieijra_2020},
finite-temperature calculations~\cite{Irikura_2020},
open quantum states~\cite{Nagy_2019,Hartmann_2019,Vincentini_2019,Yoshioka_2019},
and
quantum states with nonabelian or anyonic symmetries~\cite{Vieijra_2020}.

In the present study, as one of the intriguing targets of the RBM method, we focus on the frustrated spin systems, in which spin configurations cannot simultaneously satisfy the energy gain of the competing magnetic interactions.
The application to the frustrated spin systems is interesting and important because the QMC method 
giving numerically exact results
cannot be applied because of the sign problem. 
So far, other types of neural networks than the RBM~\cite{Cai_2018,Liang_2018,Choo_2019,Westerhout_2020,Szabo_2020}, and the combination of the RBM and Gutzwiller-projected fermion wave functions~\cite{Ferrari_2019,Nomura_arXiv} have been investigated to check the accuracy of the variational ansatz. 
However, there is no systematic study on the wave functions composed only of the RBM.
It is a fundamental problem whether one of the simplest neural networks, the RBM, can achieve high accuracy.

Here, we 
study a representative frustrated spin model,
the two-dimensional (2D) $J_1$-$J_2$ Heisenberg Hamiltonian ($J_1$: nearest-neighbor magnetic interaction, $J_2$: next-nearest-neighbor magnetic interaction) 
because almost all the previous studies based on the neural-network variational wave functions~\cite{Liang_2018,Choo_2019,Westerhout_2020,Szabo_2020,Ferrari_2019,Nomura_arXiv} have focused on the 2D $J_1$-$J_2$ Heisenberg model. 
The best accuracy of the ground-state representation achieved so far by the neural-network-only wave functions is on the order of $10^{-3}$ in terms of the relative error of the energy for the $6\times 6$ lattice (largest system size where the exact result is available) around $J_2/J_1=0.5$ (frustrated regime).

By performing systematic benchmark calculations using the RBM wave functions, we demonstrate that the RBM wave function can outperform the previously studied neural-network wave functions if it is combined with quantum-number projections. 
The quantum-number projections, which restore the symmetry of the wave function, help the RBM learn the quantum states, resulting in a drastic improvement of accuracy. 
The RBM wave function combined with the quantum-number projections achieves the accuracy of $\sim 10^{-4}$ in the relative error of the ground-state energy for the $6\times 6$ lattice around $J_2/J_1=0.5$, which is indeed one-order-of-magnitude more accurate than the existing neural-network results.
The accuracy is comparable to that obtained by other state-of-the-art wave function methods.
Furthermore, with the combined wave functions, we can also calculate the excited states with high accuracy.
The study shows a practical guideline for applying the RBM wave function to the frustrated spin systems.

The paper is organized as follows. 
In Sec.~\ref{sec_model_methods}, we introduce the 2D $J_1$-$J_2$ Heisenberg Hamiltonian and present how we prepare the RBM variational wave functions. We also discuss the way of restoring symmetry in the wave function. 
In Sec.~\ref{sec_results}, we show the benchmark results of the RBM wave functions for the $J_1$-$J_2$ Heisenberg model.
Both the ground state and excited states are investigated. 
Finally, Sec.~\ref{sec_summary} is devoted to the summary and discussion. 

\section{Model and Methods} 
\label{sec_model_methods}

\subsection{Model}

In this paper, we apply the method to the 2D $J_1$-$J_2$ Heisenberg Hamiltonian on the square lattice (we consider $L\times L$ lattices with periodic boundary condition), which reads 
\begin{eqnarray} 
 \hat{\mathcal H} =  J_1 \sum_{ \langle i, j \rangle}  {\bf S}_i  \cdot {\bf S}_j +  J_2 \sum_{ \langle \langle  i, j \rangle \rangle}  {\bf S}_i  \cdot {\bf S}_j. 
\label{J1J2Hamiltonian}
 \end{eqnarray}
Here, ${\bf S}_i$ is the spin-$1/2$ operator at site $i$, and $\langle i, j \rangle$ and $\langle \langle i, j \rangle \rangle$ denote the pair of nearest-neighbor and next-nearest-neighbor sites, respectively.
For the site specified by the coordinate ($r_x$, $r_y$), 
the nearest-neighbor sites are located at ($r_x\pm 1$, $r_y$) and ($r_x$, $r_y\pm 1$), 
and the next-nearest-neighbor sites are at  ($r_x\pm 1$, $r_y\pm 1$). 
We take $J_1$ as the energy unit ($J_1=1$) and consider positive $J_2$ (antiferromagnetic coupling).

The nearest-neighbor coupling $J_1$ favors the N\'eel-type antiferromagnetic order, whereas the next-nearest-neighbor coupling $J_2$ favors stripe-type antiferromagnetic order. 
Therefore, $J_1$ and $J_2$ compete with each other, and in the classical Heisenberg model case, there is a phase transition from the N\'eel state to the stripe state at $J_2 = 0.5$. 
If we consider the quantum fluctuation, there might emerge a quantum spin liquid phase around $J_2=0.5$. 
Although the ground-state phase diagram has been studied using various numerical wave-function methods (the QMC method cannot be applied because of the sign problem)~\cite{Jiang_2012,Hu_2013,Gong_2014,Morita_2015,Wang_2016,Wang_2018,Liu_2018,Haghshenas_2018,Ferrari_2020,Nomura_arXiv,Liu_arXiv}, 
the ground-state phase diagram around $J_2=0.5$ is still highly controversial.

As is mentioned in Sec.~\ref{Sec_intro}, for this challenging problem, various machine-learning-based methods have been applied to compare the performance. 
However, there is no systematic investigation of the accuracy of the RBM wave function. 
Here, we perform such a systematic investigation on the RBM wave function combined with quantum-number projections.

\subsection{Methods}

\subsubsection{RBM wave function}

When we apply the RBM wave function to quantum-many body systems, we first need to define a one-to-one correspondence between the states of the visible units $\sigma_i = \pm 1$ and those of the physical degrees of freedom. 
In the case of spin-1/2 Hamiltonians with the $N_{\rm site}$ spins, the most natural mapping is to prepare $N_{\rm site}$ visible units and to take $\sigma_i = 2S_i^z$. 
Then, the RBM wave function is given by 
(we do not show the normalization factor)
\begin{eqnarray}
 \Psi(\sigma)   &=&   \sum_{ \{ h_k\}}  \exp \left ( \sum_{i} a_i \sigma_i +  \sum_{i,k} W_{ik}  \sigma_i h_k +\sum_{k} b_k h_k \right ) \nonumber \\ 
 &=&  \exp \left( \sum_{i} a_i \sigma_i \right) \times  \prod_k   2 \cosh \left( b_k + \sum_{i} W_{ik}  \sigma_i  \right ),   \nonumber \\ 
 \label{eq_RBM1}
\end{eqnarray}
where $\sigma$ is the spin configuration $\sigma = (\sigma_1, \ \sigma_2, \ldots, \ \sigma_{N_{\rm site}} ) $, 
and $h_k = \pm 1$ ($k=1 ,\ldots,M$) indicates the state of the hidden units.
$a_i$ and $b_k$ are the magnetic field (called bias in the machine learning community) on the visible and hidden units, respectively, and $W_{ik} $ is the classical interaction between the visible and hidden units. 
We set the complex bias term for the visible units $a_i$ to be $a_i$ = $i \frac{\pi}{2}$ when the site $i$ belongs to one of the sublattices (``sublattice B") of the square lattice ($a_i = 0$ for the sublattice A).  
With this complex bias term $a_i = i\pi/2$ for the sublattice B, $|\!\uparrow \rangle$ ($|\!\downarrow \rangle$) state at the $i$th site
acquires a phase as follows $| \! \uparrow \rangle  \rightarrow e^{i\frac{\pi}{2} }| \! \uparrow \rangle = i | \!  \uparrow \rangle$
($| \! \downarrow \rangle  \rightarrow e^{-i\frac{\pi}{2} }| \! \downarrow \rangle = -i | \! \downarrow \rangle  $).
Physically, this corresponds to the gauge transformation in which the spin-quantization $x$ and $y$ axes are rotated by 180 degrees around the $z$ axis for the sublattice B~\cite{Carleo_2018}. 
Therefore, applying the RBM wave function in Eq. (\ref{eq_RBM1}) with the above-mentioned $a_i$ values to the Hamiltonian in Eq. (\ref{J1J2Hamiltonian}) is equivalent to applying the RBM wave function with $a_i =0$ for all the sites 
\begin{eqnarray}
\Psi(\sigma) =  \prod_k   2 \cosh \left( b_k + \sum_{i} W_{ik}  \sigma_i  \right )
\label{eq_RBM2}
\end{eqnarray}
to the following Hamiltonian with the gauge transformation being applied: 
\begin{eqnarray} 
 \hat{\mathcal H} =  J_1 \sum_{ \langle i, j \rangle}  (- S_i^x S_j^x - S_i^y S_j^y + S_i^z S_j^z )  +  J_2 \sum_{ \langle \langle  i, j \rangle \rangle}  {\bf S}_i  \cdot {\bf S}_j. \nonumber \\ 
\label{J1J2Hamiltonian2}
 \end{eqnarray}
Hereafter, we consider the RBM wave function in Eq. (\ref{eq_RBM2}) and apply it to the Hamiltonian in Eq. (\ref{J1J2Hamiltonian2}). 
The gauge transformation is helpful for mitigating the sign change in the wave function, and thus facilitating the RBM optimization. 
Indeed, when $J_2=0$, the ground-state wave function for the Hamiltonian in Eq. (\ref{J1J2Hamiltonian2})
becomes positive-definite, which allowed using real variational parameters $\{ b_k, \ W_{ik} \}$ in Ref. \cite{Carleo_2017}, which introduced the RBM wave function.
Around the frustrated regime $J_2 = 0.5$, there exist sign changes in the wave function. Therefore, we need to introduce complex $b_k$ and $W_{ik} $ variational parameters to represent the sign changes.

\begin{table}
\caption{Character table of the $C_{4v}$ point group consisting of the rotations and reflections of the square.}
\label{tab_irrep}
\footnotesize
\begin{center}
\begin{tabular}{@{\ \ \  } c  |   c @{\ \  \ \  \ }  c @{\ \ \ \ \ } c @{\ \ \ \ \ }  c @{\ \ \ \ \ }   c @{\ \ } }
\br
  &  \ \ E  & $2C_4$ & $C_2$ & $2\sigma_v$ & $2\sigma_d$  \\
\mr
$A_1$ \  & \ \  1  &  \ \  1  & \ \ 1   & \ \ 1  & \ \ 1  \\
$A_2$ \  & \ \  1  &  \ \ 1   & \ \ 1   & $-1$   & $-1$   \\
$B_1$ \  & \ \  1  & $-1$     & \ \ 1   & \ \  1 & $-1$   \\
$B_2$ \  & \ \  1  & $-1$     & \ \ 1   & $-1$   & \ \ 1  \\
$E$   \  & \ \  2  &  \ \  0  & $-2$    & \ \ 0  & \ \ 0  \\
\br
\end{tabular}
\end{center}
\end{table}

\normalsize 

\subsubsection{Projections to enforce symmetries}

The eigenstates of the Hamiltonian of finite-size systems can be classified by their symmetry. 
Because the RBM has a property of universal approximation, in the limit of a large number of hidden units, the RBM wave function can represent any eigenstates of the Hamiltonian~\cite{Clark_2018,Huang_arXiv}.
However, we need to work with a finite number of hidden units, and the optimization of variational parameters is performed numerically.  
The numerically optimized RBM wave function does not perfectly satisfy the proper symmetry.

We can restore the symmetry of the RBM wave functions using quantum-number projections~\cite{Mizusaki_2004}. 
Here, we use the combination of total-momentum, spin-parity, and lattice-symmetry projections. 
Then, we prepare four different types of wave functions as follows (throughout the section, we omit the normalization factor of the wave functions).

\begin{enumerate}
    \item  $\Psi  ( \sigma )$ in Eq. (\ref{eq_RBM2}), for which we do not apply any projections. We do not impose symmetry in the variational parameters either.  
    \item $\Psi_{\bf K}  ( \sigma )$ labelled by the total momentum $\bf K$. 
          $\Psi_{\bf K}  ( \sigma )$ is given by applying the total-momentum projection to the RBM wave function as 
          \begin{eqnarray} \ \ 
             \Psi_{\bf K}  ( \sigma ) = \sum_{{\bf R} }  e^{ - i {\bf K}  \cdot {\bf R} }     \Psi ( T_{\bf R}   \sigma ) 
              \label{Eq.mom_proj}
         \end{eqnarray}
        where $T_{\bf R}$ is a translation operator shifting all the particles by the amount ${\bf R}$.
        The wave function on the right-hand side $\Psi (\sigma)$ is the RBM wave function in Eq. (\ref{eq_RBM2}). 
        Note that the wave function on the left-hand side satisfies the symmetry, even when the wave function on the right-hand side $\Psi (\sigma)$ does not preserve the symmetry. 
    \item $\Psi_{\bf K}^{S_\pm} ( \sigma )$ labelled by the total momentum $K$ and spin-parity ($S_+$ or $S_-$): 
          \begin{eqnarray} 
               \ \  \Psi_{\bf K}^{S_\pm} ( \sigma ) = \sum_{{\bf R} }  e^{ - i {\bf K}  \cdot {\bf R} }   \left  [   \Psi ( T_{\bf R}   \sigma )  \pm  \Psi (- T_{\bf R}   \sigma )   \right ] 
               \label{Eq.spin_mom_proj}
          \end{eqnarray} 
           (double sign in the same order).
        Here, $S_+$ ($S_-$) indicates that the wave function on the left-hand side is symmetric (anti-symmetric) with respect to the global spin-flip, which means that the state resides in the even (odd) $S$ sector. 
    \item $\Psi_{\bf K}^{I,S_\pm} ( \sigma )$ labelled by the total momentum $K$, spin-parity ($S_+$ or $S_-$), and irreducible representation $I$ of the $C_{4v}$ point group (Table~\ref{tab_irrep}): 
    \begin{eqnarray}
        \Psi_{\bf K}^{I,S_\pm}  ( \sigma ) \! = \! \! \sum_{R, {\bf R} }  e^{ - i {\bf K}  \cdot {\bf R} } \chi_R^I  \left [   \Psi ( T_{\bf R} R  \sigma ) \! \pm \!  \Psi (- T_{\bf R} R   \sigma )  \right ]  \nonumber \\
        \label{Eq.full_proj} 
     \end{eqnarray}
     (double sign in the same order).
     Here, $R$ is a symmetry operation of the $C_{4v}$ point group, and $ \chi_R^I$ is the character of the $I$ representation for the symmetry operation $R$. 
\end{enumerate}
By using the combination of the quantum-number projections, one can expect that the accuracy of the wave function systematically improves~\cite{Tahara_2008}.

\subsubsection{Calculation of physical quantities and optimization of RBM wave function}

The form of the RBM wave function $\Psi(\sigma)$ in Eq. (\ref{eq_RBM2}) depends on the variational parameters $\{ b_k, W_{ik}  \}$.
Therefore, the problem of approximating eigenstate wave functions of the Hamiltonian can be recast as the optimization of the variational parameters $\{ b_k, W_{ik} \}$.
Because we are interested in the ground state or low-lying excited states (see Sec.~\ref{Sec_ex_calc} for the calculation of excited states), we optimize the RBM wave function to minimize the energy expectation value. 
The total energy $ E = \langle { \hat{\mathcal H} } \rangle =  \frac{ \langle \Psi | \hat{ \mathcal H }  | \Psi  \rangle  } { \langle \Psi | \Psi \rangle  }$ can be computed by the Monte Carlo sampling with weight $p(\sigma) \propto  | \Psi (\sigma) |^2   $ as 
\begin{eqnarray}
\label{Eq_H_expectation}
     \langle \hat{\mathcal H} \rangle  = \frac {\sum_{\sigma}     p (\sigma)  E_{\rm loc} (\sigma)} { \sum_{\sigma}     p (\sigma) }, 
\end{eqnarray}
where the local energy $ E_{\rm loc} (\sigma)$ is given by 
$ E_{\rm loc} (\sigma) =  \sum_{\sigma'}  \langle \sigma |  \hat{ \mathcal H }   | \sigma ' \rangle  \frac{  \Psi ( \sigma' ) } { \Psi(\sigma) }$. 
The energy expectation value for the variational wave functions in Eqs. (\ref{Eq.mom_proj}), (\ref{Eq.spin_mom_proj}), and (\ref{Eq.full_proj}) can be computed in the very same way. 
Note also that the expectation value for the operator $\hat{\mathcal O}$ can be calculated by replacing $\hat{\mathcal H}$ in Eq. (\ref{Eq_H_expectation}) with $\hat{\mathcal O}$.

The total energy $E$ is a highly nonlinear function with respect to the variational parameters $\{ b_k, W_{ik}  \}$. 
Therefore, the total energy can be interpreted as a loss function in the machine-learning language~\cite{Melko_2019}. 
For optimizing the variational parameters to minimize the total energy $E$, we employ the stochastic reconfiguration (SR) method~\cite{PhysRevB.64.024512}. 
The SR method can stabilize the optimization because it is equivalent to the imaginary-time Hamiltonian evolution within the Hilbert space spanned by the variational wave function. 
We refer to Ref.~\cite{Nomura_2017} for further technical details of the optimization of the RBM wave function.

 \subsubsection{Calculations of excited states}
\label{Sec_ex_calc}

With the above-described quantum-number projections, we can calculate the excited states as well as the ground state. 
The ground state is labelled by the zero total momentum (${\bf K} = {\bf 0}$), zero total spin (hence even spin parity $S_+$), and $A_1$ irreducible representation. 
For example, if we apply all the projections, the variational ground-state wave function reads   
\begin{eqnarray}
 \ \ \Psi_{\bf K = 0 }^{A_1, S_+ } ( \sigma ) = \sum_{R, {\bf R} }  \left  [   \Psi ( T_{\bf R} R  \sigma )  +  \Psi (- T_{\bf R} R   \sigma )  \right ].  
 \label{Eq.full_proj_GS}
 \end{eqnarray} 
The excited states are given by optimizing the variational wave functions for different quantum-number sectors~\cite{Choo_2018,Nomura_2020}.

\subsubsection{Calculation conditions} 
\label{sec_calc_condition}

In the present study, we apply the four types of variational wave functions in Eqs. (\ref{eq_RBM2}), (\ref{Eq.mom_proj}), (\ref{Eq.spin_mom_proj}), and (\ref{Eq.full_proj})
to the 2D $J_1$-$J_2$ Heisenberg model in Eq. (\ref{J1J2Hamiltonian2}). 
We take $L \times L$ square lattice ($N_{\rm site} = L^2$) with $L$ being an even number, and impose periodic boundary condition. 
Within this geometry, $6 \times 6$ lattice is the maximum size for which the exact solution is available. 
Therefore, we mainly focus on $6 \times 6$ lattice to take a systematic benchmark on the RBM wave functions. 
A control parameter for the accuracy of the wave functions is the number of hidden units $M$.
By increasing $M$, we can expect that accuracy improves. 

As for the variational parameters $\{b_k, W_{ik}\}$, to make it possible to express the sign change of the wave function, we take them complex numbers for $1 \leq k \leq M/2$, and the rest of the 
$b_k$ and $W_{ik}$ parameters ($M/2 + 1 \leq k \leq M$) are taken to be real (see Appendix for parameterization dependence of the RBM wave function).  
By counting the real and imaginary part of $b_k$ and $W_{ik}$ parameters as independent variational parameters, the total number of the variational parameters amounts to $3M (N_{\rm site}+1)/2$. 
As for the initial variational parameters, we put small random numbers from the interval [$-0.05$, 0.05].

The computational time scales as ${\mathcal O} (M  N_{\rm proj} N_{\rm site} )$, where $ N_{\rm proj}$ is the number of summation to apply the quantum projections in Eqs. (\ref{Eq.mom_proj}), (\ref{Eq.spin_mom_proj}), and (\ref{Eq.full_proj}). 
The $ N_{\rm proj}$ value takes $N_{\rm site}$, $2N_{\rm site}$, $16N_{\rm site}$ for  Eqs. (\ref{Eq.mom_proj}), (\ref{Eq.spin_mom_proj}), and (\ref{Eq.full_proj}), respectively [in the case of Eq. (\ref{eq_RBM2}), we do not employ the projection, hence $ N_{\rm proj}=1$].  
For the $6\times6$ lattice, when $M=36=N_{\rm site}$ and $N_{\rm proj} = N_{\rm site}$ [Eq. (\ref{Eq.mom_proj})], using 80 MPI (Message Passing Interface) parallelization on Intel Xeon 6148 CPUs, one RBM optimization iteration with 64,000 Monte Carlo measurements takes about 4 seconds. We typically perform iterations on the order of 1,000 to optimize the RBM wave functions.

\section{Results} 
\label{sec_results}
 
\subsection{Ground state}

\begin{figure}[tbp]
\vspace{0cm}
\begin{center}
\includegraphics[width=0.48\textwidth]{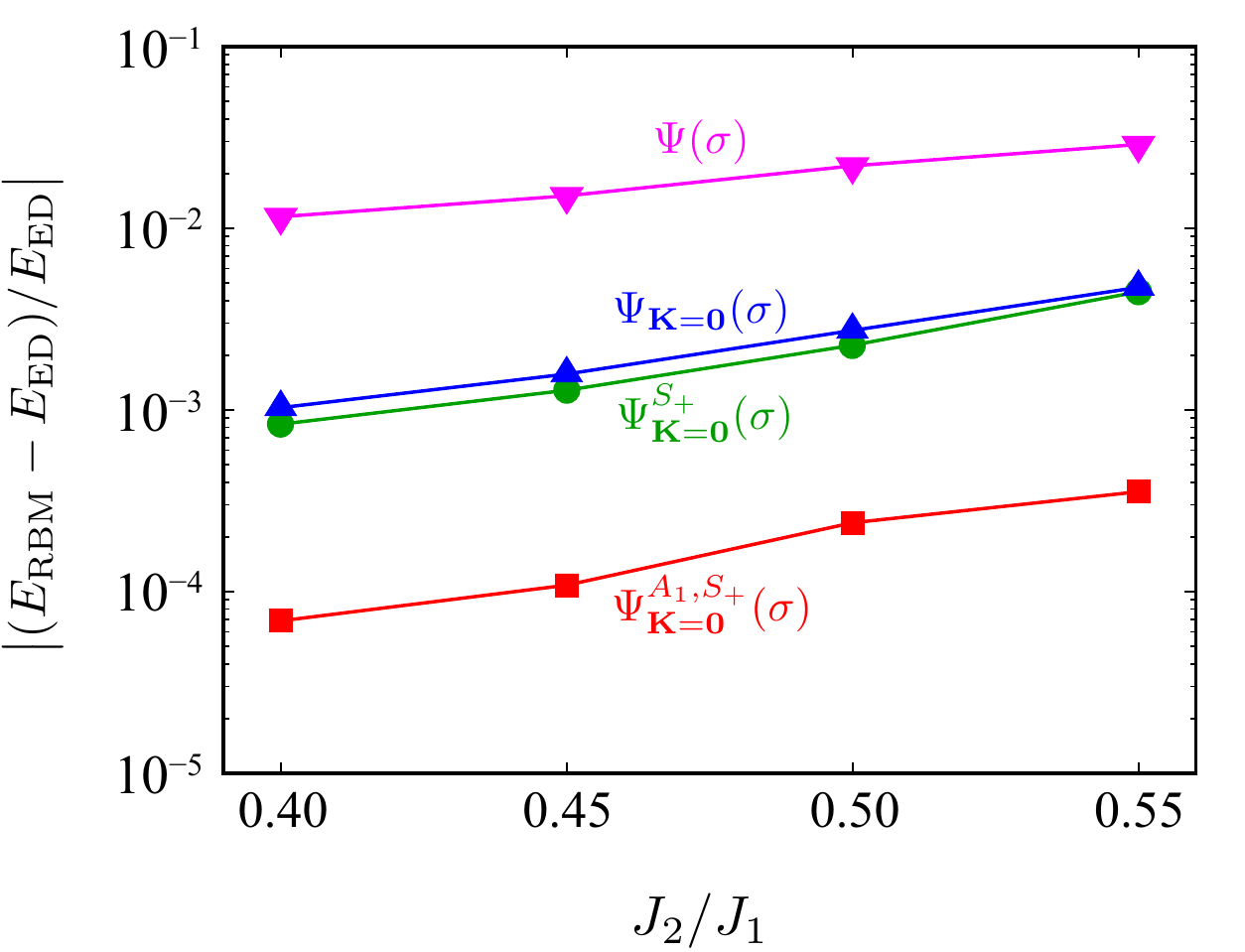}
\caption{
The relative error of the ground-state energy of the optimized RBM wave functions 
$\Psi(\sigma)$,
$\Psi_{\bf K = 0 }(\sigma)$,
$\Psi_{\bf K = 0 }^{S_+ } (\sigma)$,
and 
$\Psi_{\bf K = 0 }^{A_1, S_+ } (\sigma)$
with $M =72$ ($M$: number of hidden units)
for the $6\times6$ $J_1$-$J_2$ Heisenberg model around $J_2=0.5$.  
See Eqs. (\ref{eq_RBM2}), (\ref{Eq.mom_proj}), (\ref{Eq.spin_mom_proj}), and (\ref{Eq.full_proj}) for the difference among the four wave functions. 
The exact energy is taken from Ref.~\cite{Schulz_1996} for $J_2=0.40$, 0.50, and 0.55.
For $J_2=0.45$, the exact energy $E_{\rm ED}$ is computed using the open-source package ${\mathcal H}\Phi$~\cite{Kawamura_2017}.
}
\label{Fig_GS_symm_dep}
\end{center}
\end{figure}

We first investigate how the RBM wave function is improved by imposing symmetry. 
We apply the four different RBM wave functions 
$\Psi(\sigma)$,
$\Psi_{\bf K = 0 }(\sigma)$,
$\Psi_{\bf K = 0 }^{S_+ } (\sigma)$,
and 
$\Psi_{\bf K = 0 }^{A_1, S_+ } (\sigma)$
[see Eqs. (\ref{eq_RBM2}), (\ref{Eq.mom_proj}), (\ref{Eq.spin_mom_proj}), and (\ref{Eq.full_proj})]
for the $6\times6$ $J_1$-$J_2$ Heisenberg model around $J_2=0.5$.  
As we have already mentioned, the QMC method suffers from the negative sign problem around $J_2=0.5$, hence developing a good variational wave-function ansatz is critically important to investigate the physics of the $J_1$-$J_2$ model. 

Figure~\ref{Fig_GS_symm_dep} shows the relative error of the ground-state energy of the optimized RBM wave functions 
$\Psi(\sigma)$,
$\Psi_{\bf K = 0 }(\sigma)$,
$\Psi_{\bf K = 0 }^{S_+ } (\sigma)$,
and 
$\Psi_{\bf K = 0 }^{A_1, S_+ } (\sigma)$
with $M =72$. 
As one can see from the figure, by employing quantum-number projections, the accuracy of the RBM wave function systematically improves.  
We would like to emphasize that imposing symmetry does not correspond to preparing biased wave function such as an antiferromagnetic state, but it limits the range of high-dimensional optimization space.  
Because high-dimensional optimization is a difficult task, imposing symmetry helps the RBM to learn the ground state, leading to the improvement of the accuracy. 
The accuracy improvement can also be ascribed to the fact that the wave functions $\Psi_{\bf K = 0 }(\sigma)$,
$\Psi_{\bf K = 0 }^{S_+ } (\sigma)$,
and 
$\Psi_{\bf K = 0 }^{A_1, S_+ } (\sigma)$
can be considered as ``multi-RBM wave functions", namely a linear combination of multiple RBM wave functions. 
Then, the Hamiltonian matrix elements have off-diagonal components between two different RBMs, contributing to energy improvement.

\begin{figure}[tbp]
\vspace{0cm}
\begin{center}
\includegraphics[width=0.48\textwidth]{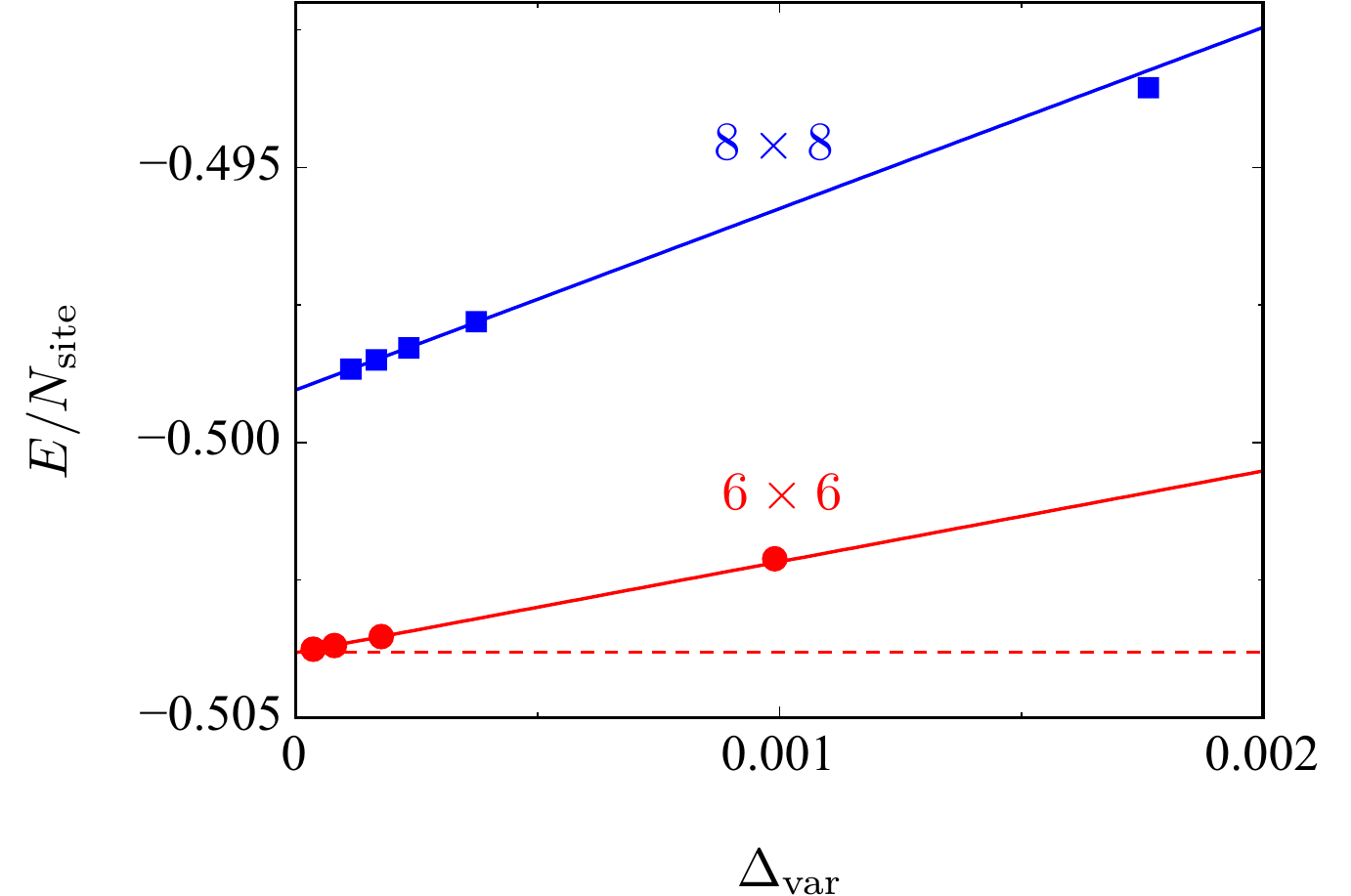}
\caption{
$M$ (number of hidden units) dependence of the ground-state energy and the dimensionless energy variance $\Delta_{\rm var} =  (\langle {\hat{\mathcal H}}^2 \rangle - \langle {\hat{\mathcal H}} \rangle^2) /  \langle {\hat{\mathcal H}} \rangle^2$ 
of the optimized $\Psi_{\bf K = 0 }^{A_1, S_+ } (\sigma)$ [Eq. (\ref{Eq.full_proj_GS})] for the $6 \times 6$ (red dots) and $8 \times 8$ (blue squares) $J_1$-$J_2$ Heisenberg model at $J_2 = 0.5$.
The data points from right to left correspond to $M= 18$, 36, 72, 144 
($M/N_{\rm site} = 0.5$, 1, 2, 4)
for the $6 \times 6$ lattice, and $M= 32$, 64, 96, 128, 192 
($M/N_{\rm site} = 0.5$, 1, 1.5, 2, 3)
for the $8\times 8$ lattice.  
By increasing $M$, both $E$ and $\Delta_{\rm var}$ decrease. 
The solid lines indicate the linear extrapolation of the total energy to the zero variance limit. 
The extrapolation is performed using the data  which satisfy $\Delta_{\rm var} < 0.0005$.
The red dashed line indicates the exact ground-state energy for the $6\times6$ lattice~\cite{Schulz_1996}.  
}
\label{Fig_GS_alpha_dep}
\end{center}
\end{figure}

In particular, the optimized RBM wave function $\Psi_{\bf K = 0 }^{A_1, S_+ } (\sigma)$ achieves the accuracy of the relative error of the energy on the order of $10^{-4}$ (0.01 \%). 
The accuracy level is quite high, and we find that the RBM wave function marks state-of-the-art accuracy compared to other neural-network-based wave functions~\cite{Choo_2019,Ferrari_2019,Nomura_arXiv} and other wave function methods~\cite{Hu_2013,Gong_2014,Morita_2015}.

So far, we have seen that symmetry helps to improve accuracy. 
The accuracy can also be improved by increasing the number of hidden units $M$. 
Figure~\ref{Fig_GS_alpha_dep} shows $M$ dependence of the ground-state energy of the optimized RBM wave function $\Psi_{\bf K = 0 }^{A_1, S_+ } (\sigma)$ for the $6\times6$ and $8\times8$ lattice at $J_2 = 0.5$.
In the case of the $6\times 6$ ($8\times8$) lattice, $M=18$, 36, 72, 144 ($M=32$, 64, 96, 128, 192) are employed. 
By increasing $M$, we see that the accuracy improves. 
The exact ground-state energy can be estimated by performing the linear extrapolation of the total energy to the zero variance limit ($\Delta_{\rm var} =  (\langle {\hat{\mathcal H}}^2 \rangle - \langle {\hat{\mathcal H}} \rangle^2) /  \langle {\hat{\mathcal H}} \rangle^2=0$)~\cite{Kwon_1998,Kashima_2001}. 
In the case of the $6\times 6$ lattice case, extrapolated energy agrees quite well with the exact ground-state energy. 
For the $8\times 8$ lattice, the extrapolation gives $E/N_{\rm site} = -0.49904(1)$. 
As a reference, we refer to the variance extrapolation using the data obtained by the VMC method combined with Lanczos steps, which gives $E/N_{\rm site} =  -0.49906(1)$ ~\cite{Hu_2013}, and the extrapolation with the DMRG truncation error, which gives  $E/N_{\rm site} = -0.4992(1)$~\cite{Gong_2014}.
Our estimate is consistent with these two references.

\begin{figure}[tbp]
\vspace{0cm}
\begin{center}
\includegraphics[width=0.4\textwidth]{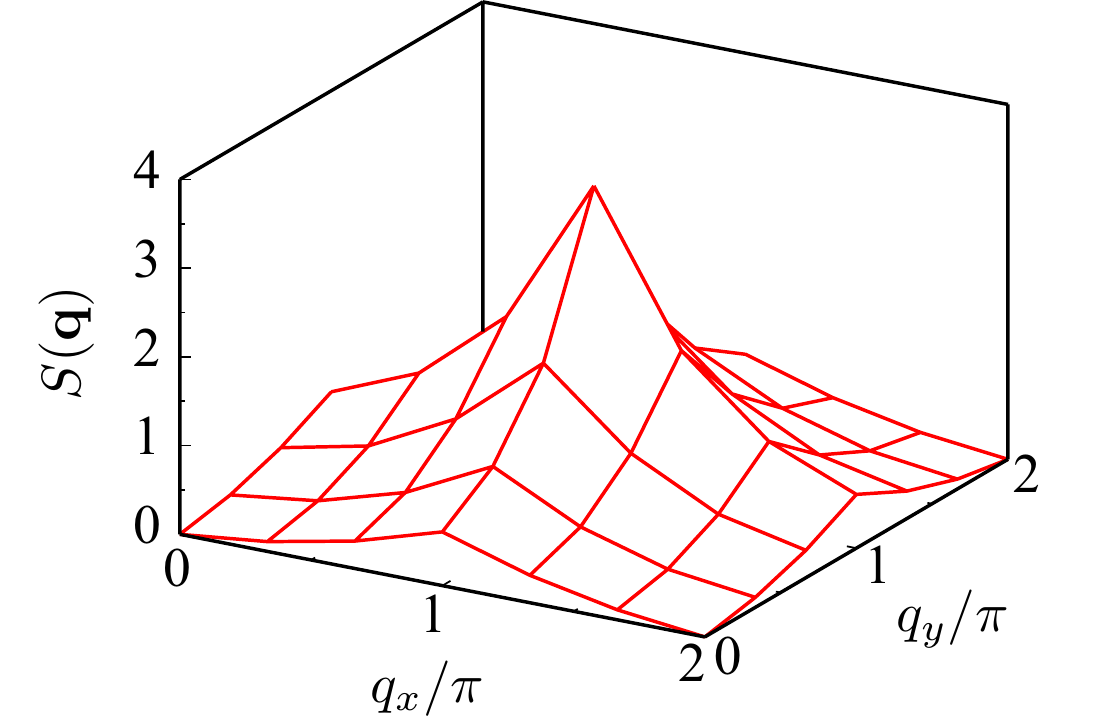}
\caption{
Spin structure factor $S ( {\bf q} ) =   \frac{1} {N_{\rm site} }  \sum_{i,j} \langle {\bf S}_i  \cdot {\bf S}_j \rangle  e ^{i {\bf q} \cdot( {\bf r}_i - {\bf r}_j )}$ for $6 \times 6$ $J_1$-$J_2$ Heisenberg model at $J_2 = 0.5$ computed with the optimized $\Psi_{\bf K = 0 }^{A_1, S_+ } (\sigma)$ [Eq.(\ref{Eq.full_proj_GS})] with $M = 144$ ($M/N_{\rm site}=4$).  
}
\label{Fig_GS_Sq}
\end{center}
\end{figure}

\begin{table}
\caption{
$M$ (number of hidden units) dependence of $S(\pi,\pi)$ for $6\times6$ $J_1$-$J_2$ Heisenberg model at $J_2 = 0.5$ calculated with the optimized $\Psi_{\bf K = 0 }^{A_1, S_+ } ( \sigma )$ [Eq.(\ref{Eq.full_proj_GS})].
The exact $S(\pi,\pi)$ value is 3.50856, which is obtained using the open-source package ${\mathcal H}\Phi$~\cite{Kawamura_2017}. 
}
\label{tab_m2}
\footnotesize
\begin{center}
\begin{tabular}{ @ {\ \ }  c @{\ \  \ \  \ }  c @{\ \ \ \ \ } c @{\ \ \ \ \ }  c @{\ \ \ \ \ }  c @{\ \ } }
\br
 $M= 18$  & $M=36$ & $M=72$ & $M=144$ & Exact   \\
\mr
 3.645(2)  & 3.539(2) & 3.521(2) & 3.505(2) & 3.50856 \\
\br
\end{tabular}
\end{center}
\end{table}

With the optimized wave functions, we can obtain not only the total energy but also other physical quantities such as the correlation function. 
Here, we compute the spin-spin correlation $\langle {\bf S}_i  \cdot {\bf S}_j \rangle$ with the optimized $\Psi_{\bf K = 0 }^{A_1, S_+ } ( \sigma )$ and obtain the spin structure factor $S ( {\bf q} ) =   \frac{1} {N_{\rm site} }  \sum_{i,j} \langle {\bf S}_i  \cdot {\bf S}_j \rangle  e ^{i {\bf q} \cdot( {\bf r}_i - {\bf r}_j )}$ for the $6 \times 6$ lattice at $J_2 = 0.5$. 
At $J_2 = 0.5$, the spin structure factor $S ( {\bf q} )$ has a peak at ${\bf q}  = (\pi, \pi)$ (Fig.~\ref{Fig_GS_Sq}).
Table~\ref{tab_m2} shows the peak value of the structure factor $S(\pi,\pi)$ for the four different $M$ cases. 
By increasing the number of hidden units $M$, not only the accuracy of the energy but also that of the correlation function improves. 
When $M=144$, we see a very good agreement between the RBM and exact results for all the {\bf q} points.

\subsection{Excited States}

As described in Sec.~\ref{Sec_ex_calc}, the excited states can be calculated by optimizing the RBM wave function in quantum number sectors different from that of the ground state.
The information of excited states is essential in understanding the nature of the phases. 
In particular, if there is a phase transition, the excited-state character will change. 
Hence, the change in excited-state character can be a signature for the phase transition. 
Indeed, for example, in the case of the $J_1$-$J_2$ Heisenberg chain, the level crossing of singlet and triplet excited states at finite size systems and extrapolation to the thermodynamic limit give an elegant estimate for the fluid-dimer phase transition~\cite{Okamoto_1992}. 
Recently, a similar level spectroscopy has also been performed in the 2D $J_1$-$J_2$ Heisenberg model~\cite{Wang_2018,Ferrari_2020,Nomura_arXiv}. 
There, the level crossing between the singlet $S=0$ excited state with total momentum ${\bf K}= (\pi,0)$ and the triplet $S=1$ excited state with total momentum ${\bf K}= (\pi,\pi)$ is highlighted as a possible hallmark for the phase transition between quantum spin liquid and valence bond solid phases~\cite{Nomura_arXiv}.

\begin{table}
\caption{Character table of the $C_{2v}$ point group.}
\label{tab_irrep2}
\footnotesize
\begin{center}
\begin{tabular}{@{\ \ \  } c  |   c @{\ \  \ \  \ }  c @{\ \ \  }  c @{\ \ \ }   c @{\ \ } }
\br
  &  \ \ E  & $C_2$ & $\sigma_v(xz)$ & $\sigma_v(yz)$  \\
\mr
$A_1$ \  & \ \  1  &  \ \  1  & \ \ 1   & \ \ 1   \\
$A_2$ \  & \ \  1  &  \ \ 1   & $-1$   & $-1$    \\
$B_1$ \  & \ \  1  & $-1$     & \ \ 1   & $-1$   \\
$B_2$ \  & \ \  1  & $-1$     & $-1$   & \ \ 1    \\
\br
\end{tabular}
\end{center}
\end{table}

\begin{figure}[tbp]
\vspace{0cm}
\begin{center}
\includegraphics[width=0.48\textwidth]{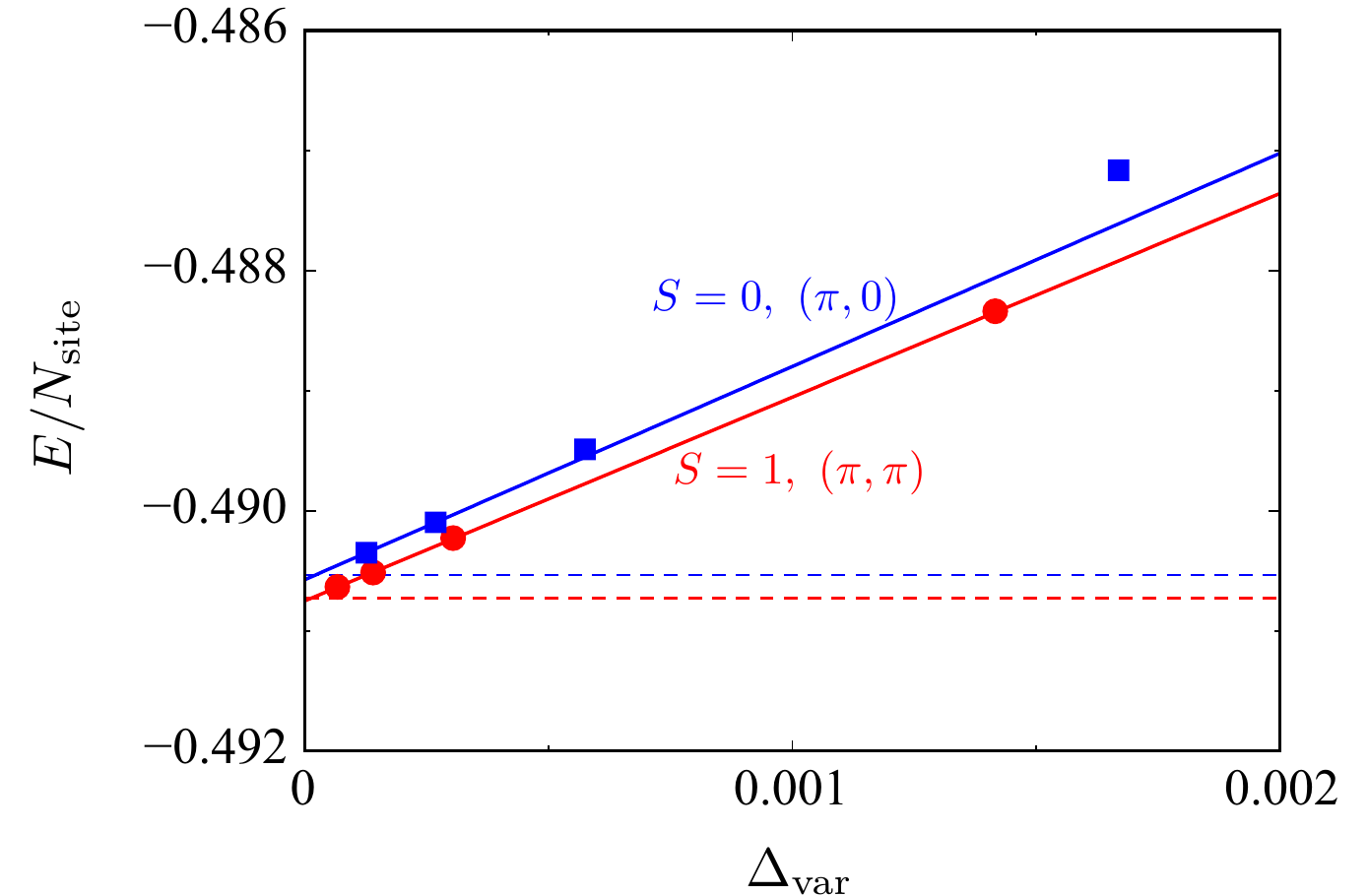}
\caption{
$M$ (number of hidden units) dependence of the excited-state energy and the dimensionless energy variance $\Delta_{\rm var} =  (\langle {\hat{\mathcal H}}^2 \rangle - \langle {\hat{\mathcal H}} \rangle^2) /  \langle {\hat{\mathcal H}} \rangle^2$ 
of the optimized $\Psi_{\bf K = (\pi,\pi)}^{A_1, S_- } (\sigma)$ [$S=1$ excited state with ${\bf K}=(\pi,\pi)$, red dots] and $\Psi_{ {\bf K} = (\pi,0)}^{B_1, S_+ } (\sigma)$ [$S=0$ excited state with ${\bf K}=(\pi,0)$, blue squares] for the $6 \times 6$ $J_1$-$J_2$ Heisenberg model at $J_2 = 0.5$.
The data points from right to left correspond to $M= 18$, 36, 72, 144. 
By increasing $M$, both $E$ and $\Delta_{\rm var}$ decrease. 
The solid lines indicate the linear extrapolation of the total energy to the zero variance limit. 
The extrapolation is performed using the data  which satisfy $\Delta_{\rm var} < 0.0005$.
The dashed lines indicate the exact excited-state energies calculated using ${\mathcal H}\Phi$~\cite{Kawamura_2017}. 
}
\label{Fig_EX_alpha_dep}
\end{center}
\end{figure}

Here, we focus on the same level crossing. 
To compute the triplet $S=1$ excited state with total momentum ${\bf K}= (\pi,\pi)$ [singlet $S=0$ excited state with total momentum ${\bf K}= (\pi,0)$], we optimize the RBM wave function in the odd total-spin ($S_-$) and ${\bf K}= (\pi,\pi)$ sector [even total-spin ($S_+$) and ${\bf K}= (\pi,0)$ sector]. 
Because the $S=1$ and $S=0$ excited states are the lowest energy state in odd ($S_-$) and even ($S_+$) total-spin sector, respectively, the optimizations in odd ($S_-$) and even ($S_+$) total-spin sector result in the $S=1$ and $S=0$ excited states, respectively. 
We have confirmed that we indeed obtain $S=1$ and $S=0$ excited states by directly computing the expectation value of the total spin with the optimized wave functions. 
We also use the point-group projection. 
The lowest energy state in $S=1$, ${\bf K}= (\pi,\pi)$ sector is characterized by the $A_1$ representation of the $C_{4v}$ point group. 
In the case of $S=0$, ${\bf K}= (\pi,0)$ sector, we need to refer to the character table of $C_{2v}$ point group (Table~\ref{tab_irrep2}), and the lowest energy state has a property of the $B_1$ representation.

Figure~\ref{Fig_EX_alpha_dep} shows the $M$ dependence of the excited-state energy of the optimized RBM wave functions $\Psi_{\bf K = (\pi,\pi)}^{A_1, S_- } (\sigma)$ [$S=1$ excited state with ${\bf K}=(\pi,\pi)$, red dots] and $\Psi_{{\bf K} = (\pi,0)}^{B_1, S_+ } (\sigma)$ [$S=0$ excited state with ${\bf K}=(\pi,0)$, blue squares] for the $6\times6$ lattice at $J_2 = 0.5$.
By increasing $M$ as $M = 18$, 36, 72, 144, the energy improves.  
As in the case of the ground state (Fig.~\ref{Fig_GS_alpha_dep}), the variance extrapolation gives an excellent estimate of the exact excited-state energies.

\begin{figure}[tbp]
\vspace{0cm}
\begin{center}
\includegraphics[width=0.45\textwidth]{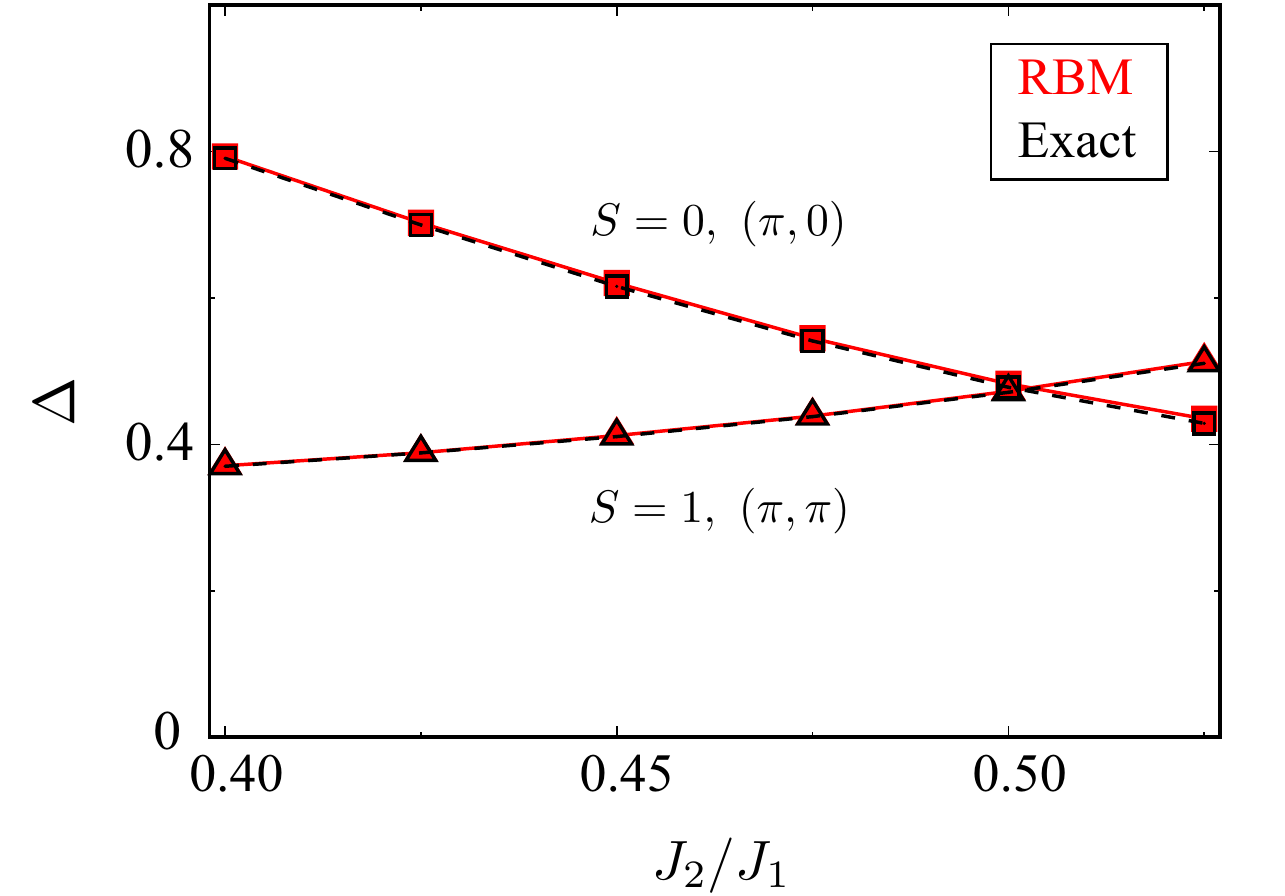}
\caption{
$J_2$ dependence of the excited energies of 
$S=0$ excited state with total momentum ${\bf K}=(\pi,0)$ (squares) and $S=1$ excited state with total momentum ${\bf K}=(\pi,\pi)$ (triangles)
for the $6\times 6$ lattice. 
The RBM data are calculated by the energy difference between the optimized $\Psi_{\bf K = (\pi,\pi)}^{A_1, S_- } (\sigma)$ and $\Psi_{\bf K = {\bf 0}}^{A_1, S_+ } (\sigma)$, 
and 
$\Psi_{{\bf K} = (\pi,0)}^{B_1, S_+ } (\sigma)$ and $\Psi_{\bf K = {\bf 0}}^{A_1, S_+ } (\sigma)$, respectively. 
The number of hidden units $M$ is 144. 
The exact excited-state energies are obtained by  ${\mathcal H}\Phi$~\cite{Kawamura_2017}.
}
\label{Fig_EX_J2_dep}
\end{center}
\end{figure}

Finally, Figure~\ref{Fig_EX_J2_dep} shows the $J_2$ dependence of the excited energies of 
singlet excited state with ${\bf K}=(\pi,0)$ (squares) and triplet excited state with ${\bf K}=(\pi,\pi)$ (triangles) for the $6\times 6$ lattice.  
The excited energies computed by the optimized RBM wave functions [$\Psi_{\bf K = (\pi,\pi)}^{A_1, S_- } (\sigma)$, $\Psi_{ {\bf K} = (\pi,0)}^{B_1, S_+ } (\sigma)$, and $\Psi_{\bf K = {\bf 0}}^{A_1, S_+ } (\sigma)$] show a very good agreement with the exact results. 
Around $J_2=0.5$, the excited energies of the singlet and triplet excitations show the level crossing. 
We leave the investigation of the size dependence of the level crossing as an interesting future problem. 

\section{Summary and Discussion} 
\label{sec_summary}

We have applied the RBM wave functions with different symmetrization levels to the 2D $J_1$-$J_2$ Heisenberg model. 
We have seen that by applying the quantum-number projections to the RBM wave function and controlling the number of hidden units, we can achieve state-of-the-art accuracy not only in the ground-state calculation but also in the excited-state calculations.
It is remarkable that the RBM, which is one of the simplest neural networks with only one hidden layer, can outperform other (deep) neural networks employed, e.g., in Ref.~\cite{Choo_2019}, with the help of symmetry. 
We emphasize again that restoring symmetry does not correspond to preparing biased wave function, but it makes the optimization space smaller and facilitates the optimization. 
We believe that the symmetry consideration will also be helpful when we apply the RBM wave function to other frustrated spin systems.

The number of variational parameters $N_{\rm var}$ ($=3M (N_{\rm site}+1)/2$, see Sec.~\ref{sec_calc_condition}) in the wave function $\Psi_{\bf K = {\bf 0}}^{A_1, S_+ } (\sigma)$ with $M=72$ for the $6\times 6$ lattice used in Fig.~\ref{Fig_GS_symm_dep} is 3996. 
This number is still tractable and small, but the $N_{\rm var}$ value to achieve the same accuracy level will grow at least as $N_{\rm var} \sim N_{\rm site}^2$ when we increase the system size.  
This is because, from the comparison between $6\times6$ and $8\times8$ results in Fig.~\ref{Fig_GS_alpha_dep}, it can be inferred that the number of hidden units $M$ to achieve the same accuracy level for different system sizes scales at least linearly with respect to $N_{\rm site}$. 
Then, for example, for $12\times 12$ lattice with four times large $N_{\rm site}$, $N_{\rm var}$ on the order of 100,000 might be required to achieve the accuracy. 
In general, a larger $N_{\rm var}$ makes optimization more difficult; therefore, it is helpful to reduce $N_{\rm var}$ value if we want to investigate large system sizes. 
One possible way to reduce $N_{\rm var}$ is to combine the RBM with other powerful wave functions.
The combination of the RBM and the Gutzwiller-projected fermion wave function is one of the examples of such attempts~\cite{Ferrari_2019,Nomura_arXiv}.
It is an important future issue to investigate the efficiency of the optimization of the RBM wave functions in large system sizes and compare it with other methods.

\section*{Acknowledgments}

We would like to thank Masatoshi Imada for fruitful discussions.
We acknowledge the financial support by Grant-in-Aids for Scientific Research (JSPS KAKENHI) (Grants No. 16H06345, No. 17K14336, No. 18H01158, and No. 20K14423).
This work was supported by MEXT as ``Program for Promoting Researches on the Supercomputer Fugaku" (Basic Science for Emergence and Functionality in Quantum Matter).
A part of the calculations was performed at Supercomputer Center, Institute for Solid State Physics, University of Tokyo.

\section*{Appendix}

As we described in Sec.~\ref{sec_calc_condition}, in the present study, the half of variational parameters $\{b_k, W_{ik}\}$ are taken to be complex,
and the rest of the $b_k$ and $W_{ik}$ parameters are taken to be real.  
This choice was rather empirically introduced. The real RBM can only control the amplitude of the wave function, while the complex RBM can also represent the phase of the wave function. Our na\"ive guess was that, by combining the real RBM and the complex RBM, the former is more sensitive to the amplitude part and the latter to the phase part. 

\begin{figure}[tbp]
\vspace{0cm}
\begin{center}
\includegraphics[width=0.46\textwidth]{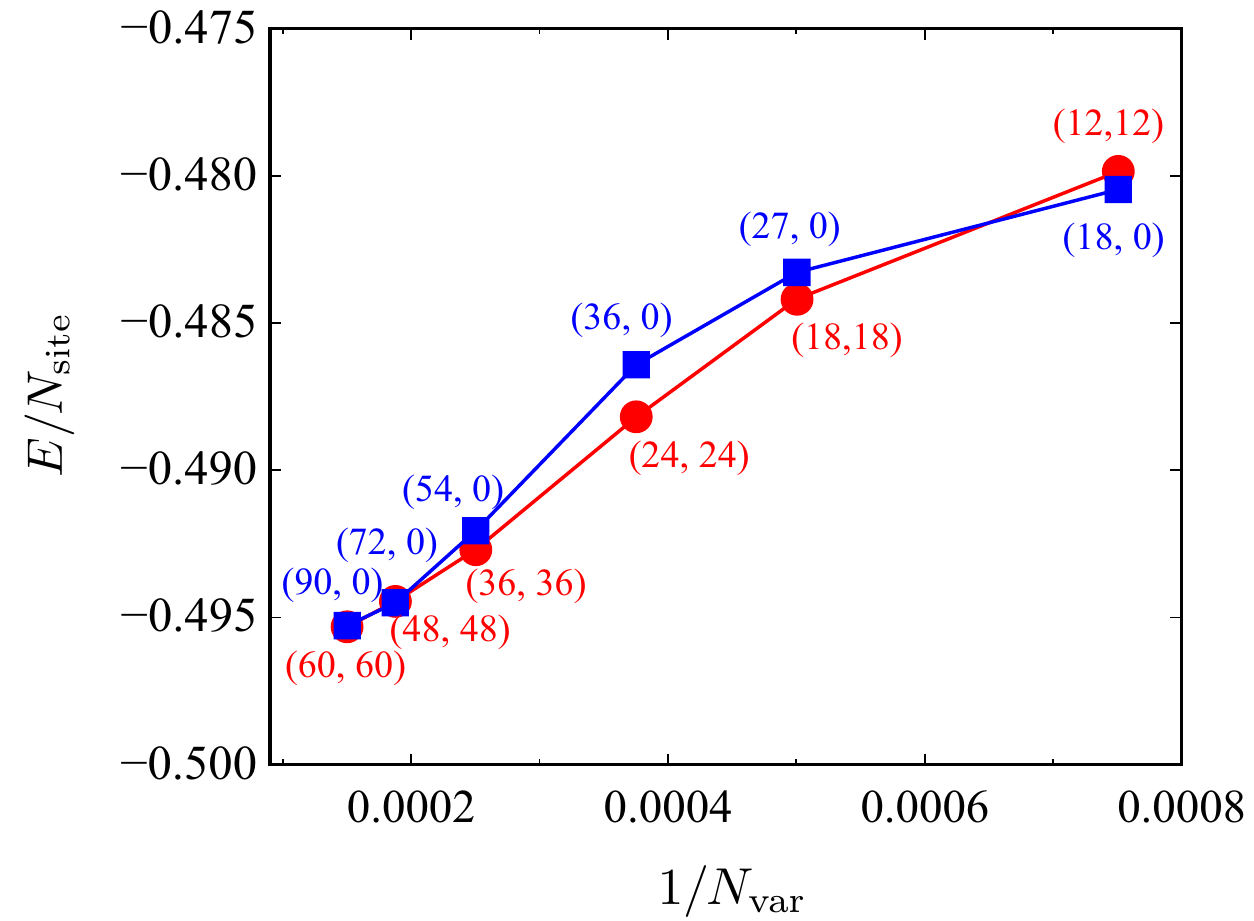}
\caption{
The comparison of the ground-state energy between the all-complex RBM (blue squares) and half-real and half-complex RBM wave function (red dots) for the $6 \times 6$ lattice at $J_2=0.5$. 
We do not utilize the quantum-number projections. 
The values in the parentheses show the number of hidden units having complex parameters (left) and that with real parameters (right).
$N_{\rm var}$ is the number of variational parameters. 
}
\label{Fig_different_RBMs}
\end{center}
\end{figure}

To see the effect of different choices of real- and complex-RBM combination, we have performed benchmark calculations for the $6 \times 6$ lattice at $J_2=0.5$. 
We prepare the RBM wave function without projections $\Psi(\sigma)$ with the setting employed in the paper (the half of hidden units have real parameters, and the rest have complex parameters) and compare it with that with all parameters being complex (Fig.~\ref{Fig_different_RBMs}). 
When the number of variational parameters is small, the all-complex RBM wave function shows a better performance, but, by increasing the variational parameters, the RBM employed in the present paper shows better accuracy. Finally, when the number of variational parameters becomes large, both wave functions show similar accuracy. 

Although the complex RBM wave function has a property of universal approximation, which ensures that the RBM can exactly represent the eigenstates of the many-body systems, the RBM with a finite number of parameters has a limit in representability. 
The result indicates that, when the number of parameters is small, the representability depends on how we design the architecture of the RBM network (we do not know the best architecture of the RBM and finding the best choice is beyond the scope of the present study). 
However, at the same time, when the number of parameters becomes large, the different parameterization of the RBM wave function does not affect the result, and the different RBMs can be systematically improved similarly. 
Indeed, as we see in Figs.~\ref{Fig_GS_alpha_dep} and \ref{Fig_EX_alpha_dep}, the accuracy of the half-real and half-complex RBM wave function systematically improves toward the exact solutions both for the ground and excited states.

\section*{References}

\bibliographystyle{iopart-num}
\bibliography{main}

\end{document}